# Impact of reaction temperatures on the particle size of $V_2O_5$ synthesized by facile hydrothermal technique and their auspicious photocatalytic performance in dye degradation

**M.A. Jalil [1], M.N.I. Khan[2], S. Mandal[3], F.-U.-Z. Chowdhury[1], M.M. Hossain[1], D. Jana[3], M.S. Alam[4], and M.M. Uddin[1*]**

[1]Department of Physics, Chittagong University of Engineering and Technology (CUET), Chattogram 4349, Bangladesh
[2]Materials Science Division, Atomic Energy Center, Dhaka 1000, Bangladesh.
[3]Department of Physics, University of Calcutta, 92 A P C Road, Kolkata-700009, West Bengal, India
[4]Department of Physics, Faculty of Science, University of Chittagong, Chattogram, 4331, Bangladesh
E-mail: mohi@cuet.ac.bd (Md. Mohi Uddin)

## ABSTRACT

In this study, a complete study of the effect of hydrothermal reaction temperatures on the synthesis and physical properties of $V_2O_5$ using the green facile mild hydrothermal method has been performed with six different temperatures 100 °C to 200 °C, in the step of 20 °C. . The XRD pattern confirm the stable orthorhombic crystal structure of the synthesized samples at all reaction temperatures. The SEM and TEM images demonstrate the particle-like morphology, and these characterizations affirmed that the particles size became larger with the increase of reaction temperatures. The FTIR analysis is employed to study the functional groups, and the obtained results are consistent with the XRD analysis. The bandgap has been estimated at various reaction temperatures using UV-vis diffuse reflectance spectra (UV-DRS) and was found to be varied 2.09 eV to 2.15 eV that are suitable range to absorb a significant amount of visible light. The photocatalysis of methylene blue (MB) with synthesized samples has been accomplished to investigate photocatalytic efficiency. The pure $V_2O_5$ synthesized at lower reaction temperature (100 °C) possess a lower bandgap and, accordingly, higher photocatalytic efficiency.

## 1. INTRODUCTION

The transition metal oxides (TMOs) is a unique materials family that scientists spent a great effort to find their fascinating applications [1,2]. The Vanadium Oxides (VOs) have drawn significant interest as a potential candidate in the TMOs family among researchers due to low-cost facile synthesis and attractive surface morphology [3,4]. Furthermore, various oxidation states (+2, +3, +4, and +5) of VOs expose a vast research field to the scientists[5]. Accordingly, the vanadium compounds form VO, $V_2O_3$, $VO_2$, and $V_2O_5$. Among these, vanadium pentoxide ($V_2O_5$) has been studied rigorously over the last decade as it has interesting and charming physical and chemical properties [6]. However, $V_2O_5$ is a promising candidate for photocatalysis [7–9], electrochemistry [10], chemical sensor [11,12], battery materials [13,14], etc. Recently, the pure and doped $V_2O_5$ was studied extensively for photocatalytic and environmental purification processes due to its lower bandgap and attractive surface morphology [9,15]

There are several dry and wet processes accessible for the synthesis of $V_2O_5$, for instant, sol-gel [15,16], decomposition [17], sputtering[18], thermal evaporation method [19], and hydrothermal method [20,21] and so on. The mild hydrothermal method is one of them that is suitable for synthesizing transition metal oxides [22] Where reaction temperatures are maintained from 100 °C to 200 °C. The numerous research articles have been published on the synthesis of vanadium compounds using the hydrothermal method (HTM). The HTM is efficient for the synthesis of $V_2O_5$ to achieve improved properties of $V_2O_5$ with different structure and morphology by varying different parameters during the synthesis process. For instant, Juyi Mu et al. and Young Bok Kim et al. have found that acid and solvent combinations significantly impact the morphology of $V_2O_5$; their reaction time and temperature were constant. They show that the belt, flowers, balls, wires and rods shaped morphology can be found merely by changing the combination of acid and solvent [3,23]. Devika P. Nair et al. and M. Zeng et al. studied the effect of surfactants on the properties of $V_2O_5$ using the HTM [24,25]. The ammonium metavanadate ($NH_4VO_3$) is a commonly used precursor material as the source of vanadium in the HTM. Several articles are reported on the synthesis of $V_2O_5$ using the $NH_4VO_3$ with different reaction temperatures at 180°C [3,9,24,26–29], 120°C [30], 160°C [31]. Furthermore, W. Yu et al. synthesized $V_2O_5$ at both 140°C and 180°C reaction temperatures [32]. All researchers have been used the $NH_4VO_3$ as precursor material. To the best of our knowledge, there is no systematic complete study on the surface morphology and physical properties of $V_2O_5$ using $NH_4VO_3$ by varying hydrothermal low to high reaction temperatures. We have successfully synthesized single-phase $V_2O_5$ starting from reaction temperature 100°C to 200°C in the step of 20°C. We have studied the structural, morphological, optical properties of the synthesized samples. The photocatalytic properties of the synthesized samples has also been tested under visible light with methylene blue (MB).

## 2. EXPERIMENTAL DETAILS

## 2.1. Materials

In this works, ammonium meta-vanadate ($NH_4VO_3$) with purity ≥ 99% (Sigma-Aldrich) was used as precursor. Both ethanol and Nitric acid were parched from Mark, Germany. All chemical were used without further purification.

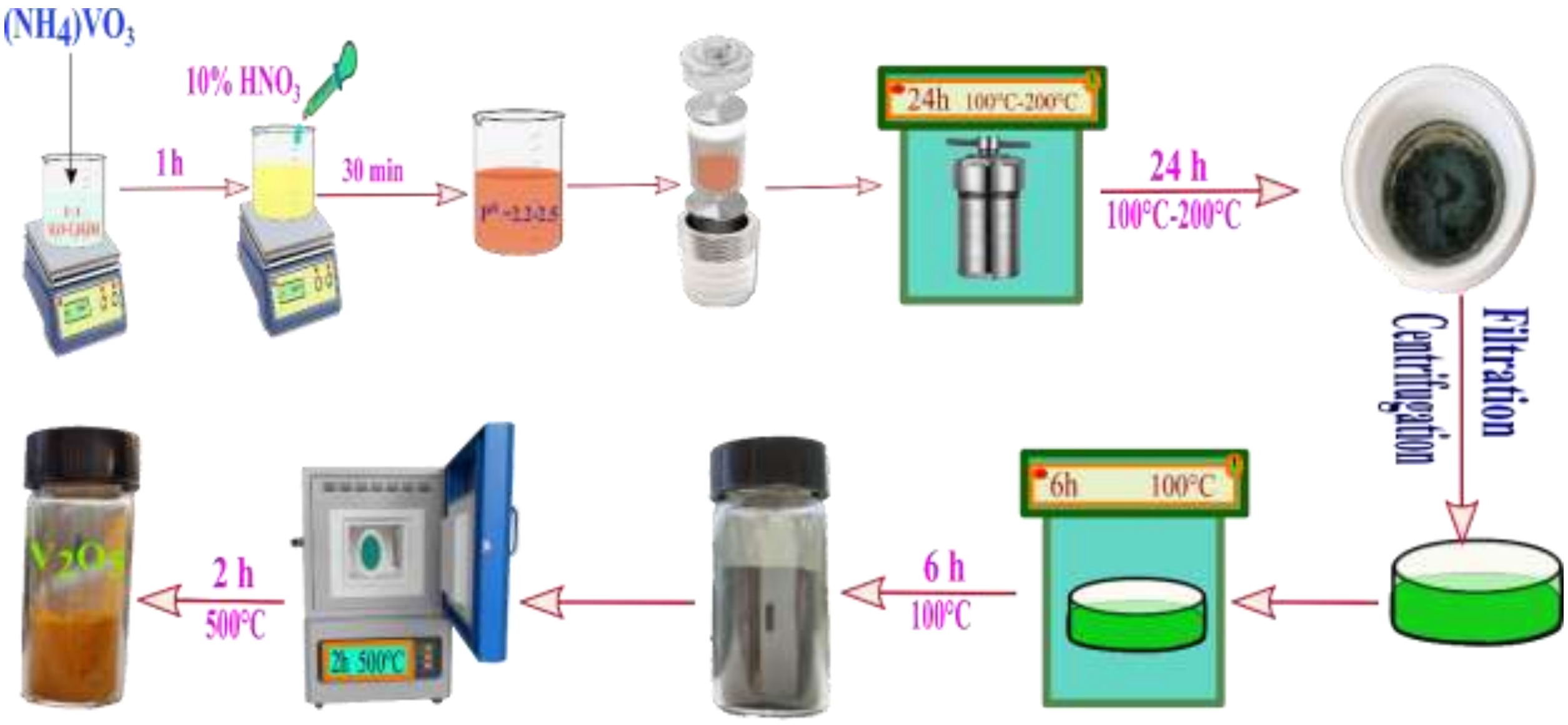

Fig. 1 The schematic illustration of facile mild hydrothermal process of sample preparation.

## 2.1. Synthesis of $V_2O_5$

In this present study, the synthesis of pure $V_2O_5$ was carried out using a facile mild hydrothermal method (FMHTM) at different reaction temperatures (100°C-200°C, in the step of 20°C). All steps involved in the synthesis procedure are shown schematically in figure 1. First, 2.3396 gm (1M) of ammonium metavanadate was dissolved in 1:1 DI water and ethanol solvent. This solution was continuously stirred for 1h at room temperature. Then freshly prepared 10% $HNO_3$ was added to the light yellow solution to maintain the pH 2.2-2.5. After that, the dark wine colour solution was transferred to a Teflon-lined stainless steel autoclave (250 ml) and heated at the desired reaction temperature in an oven for 24 hrs. We prepared six samples series with reaction temperature 100°C to 200°C in the step of 20°C. After cooling to room temperature, the obtained produced washed several times by DI water and ethanol using filtration and centrifugation. Then the green color gel was dried for 6 hrs at 100°C. Finally, powder $V_2O_5$ was found, followed by 600°C calcination in a high precision furnace for 2 hrs.

## 2.2. Characterization Techniques

The structural analysis was performed by the X-ray diffraction spectra (XRD) of the synthesized samples with the Rigaku Smart Lab diffractometer (Cu-Kα radiation and λ=1.5406 Å). The scanning electron microscopy (SEM) (JEOL, JSM6010LA), and the transmission electron microscopy (TEM)(TALOS F200 G2, Thermo Fisher Scientific, USA) were employed to study surface morphology and the particle size distribution. A Fourier-transfer infrared spectrophotometer (FT-IR) (Jupiter, STA449 F3) was used to analyze the vibrational modes of the synthesized samples. The bandgap was also estimated using UV-vis diffuse reflectance spectra (UV-DRS) (LAMBDA 1050, PerkinElmer-USA).

## 2.3. Photocatalytic Efficiency Test

The Photocatalytic efficiency test was performed in the methylene blue (MB) dye using the synthesized samples as a catalyst, and the procedure is given below:-

First, the aqueous solution of MB dye (1 x $10^{-5}$ M, 25 mL) was prepared by vigorous stirring for 30 min to get a homogeneous solution and the solvent was deionized water (DI) water. Then, 5 mg of the catalyst was added to the solution and again magnetically stirred in the dark environment for 30 min to reach the adsorption equilibrium of dye by the catalyst. The efficiency analyzing process was commenced by irradiating the solution with visible light.  The degradation of MB was continued for 180 minutes, and the absorbance was taken in 20 minutes intervals. In addition, the catalyst material was separated from the testing solution by centrifugation for 10 minutes before putting the solution in the UV-vis spectrometer.

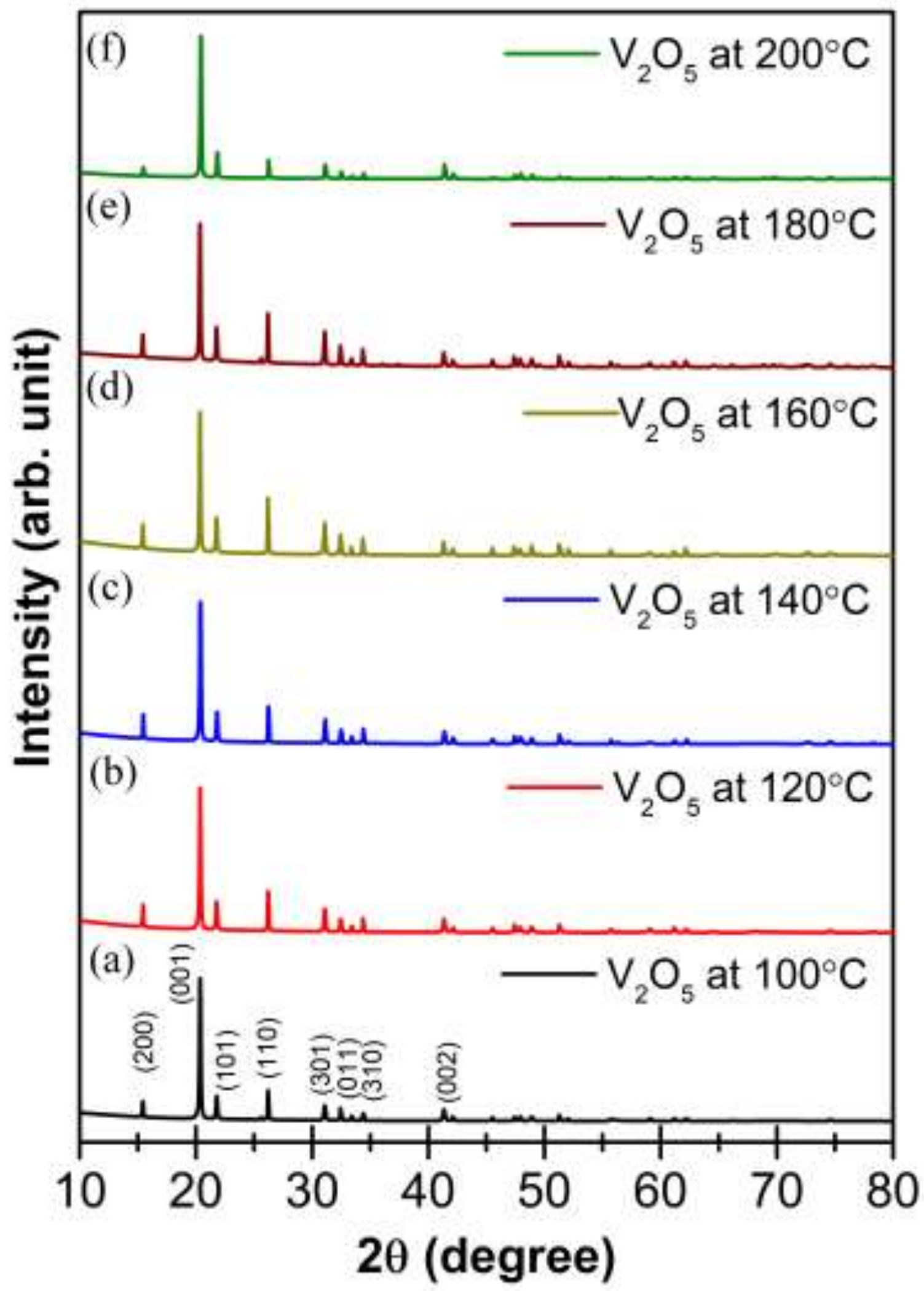

Fig.2 The XRD pattern of $V_2O_5$ synthesized by facile mild hydrothermal method (FMHTM) at different reaction temperatures (a) 100 °C, (b) 120 °C, (c) 140 °C, (d) 160 °C, (e) 180 °C, and (f) 200 °C

# 3. RESULTS AND DISCUSSION

## 3.1. Crystal Structure Analysis

The crystalline structure of as-synthesized samples were studied using the XRD measurement. Figure 2 shows the XRD patterns of powder samples synthesized at different hydrothermal reaction temperatures. It is seen that the most substantial peaks are typical in all samples, and these are (200), (001), (101), (110), (301), (011), (310) and (002) that indicates all samples have orthorhombic crystal structure with the space group *Pmmn* (JCPDS #09-0387)[33]. The most intense peak is (001) at 2θ value of 20.4°indicating the crystallites are preferably oriented at ⟨0 0 1⟩ direction. However, there are no other additional peaks found in the XRD analysis of all samples (Fig. 2). Therefore, it can be summarized from the crystal structure analysis that high purity $V_2O_5$ powder without impurities have been synthesized using FMHTM at different reaction temperatures.

## 3.2. Morphological Analysis

The morphologies of the as-synthesized $V_2O_5$ have been analyzed at room temperature by varying reaction temperatures using a scanning electron microscope (SEM) as illustrated in Figure 3 (a-f).

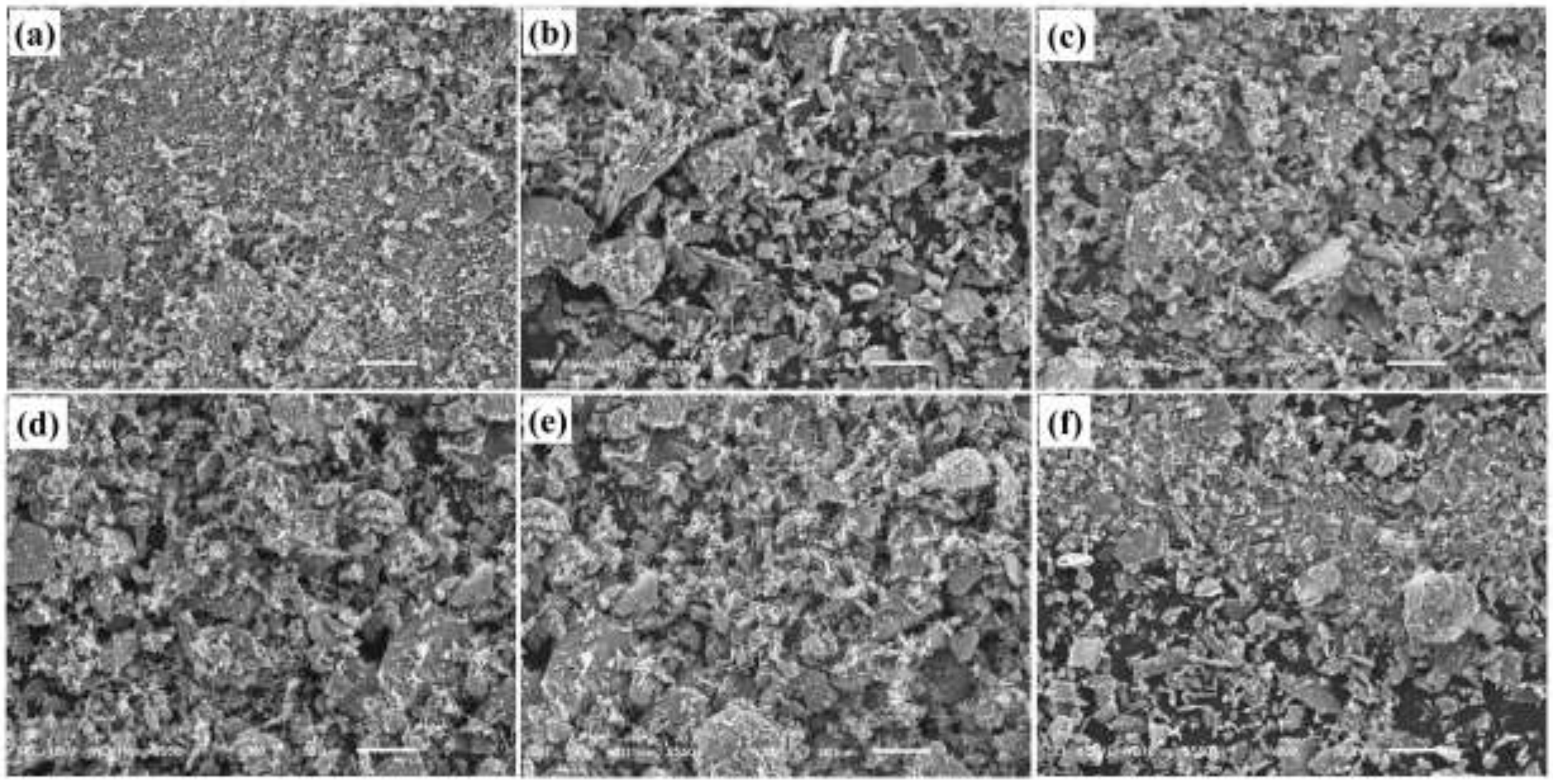

Fig-3: SEM images of $V_2O_5$ measured at room temperature with different reaction temperatures (a) 100 °C, (b) 120 °C, (c) 140 °C, (d) 160 °C, (e) 180 °C, and (f) 200 °C.

It is seen from the images that all samples alike and agglomerated particles exhibited surface morphology in all reaction temperatures. It is obvious from the images that the agglomeration increases with increasing reaction temperature. This finding is consistent with basic mechanism of agglomeration, where the surface free energy reduces due to increasing particles size and decreasing surface area results the agglomeration takes place.

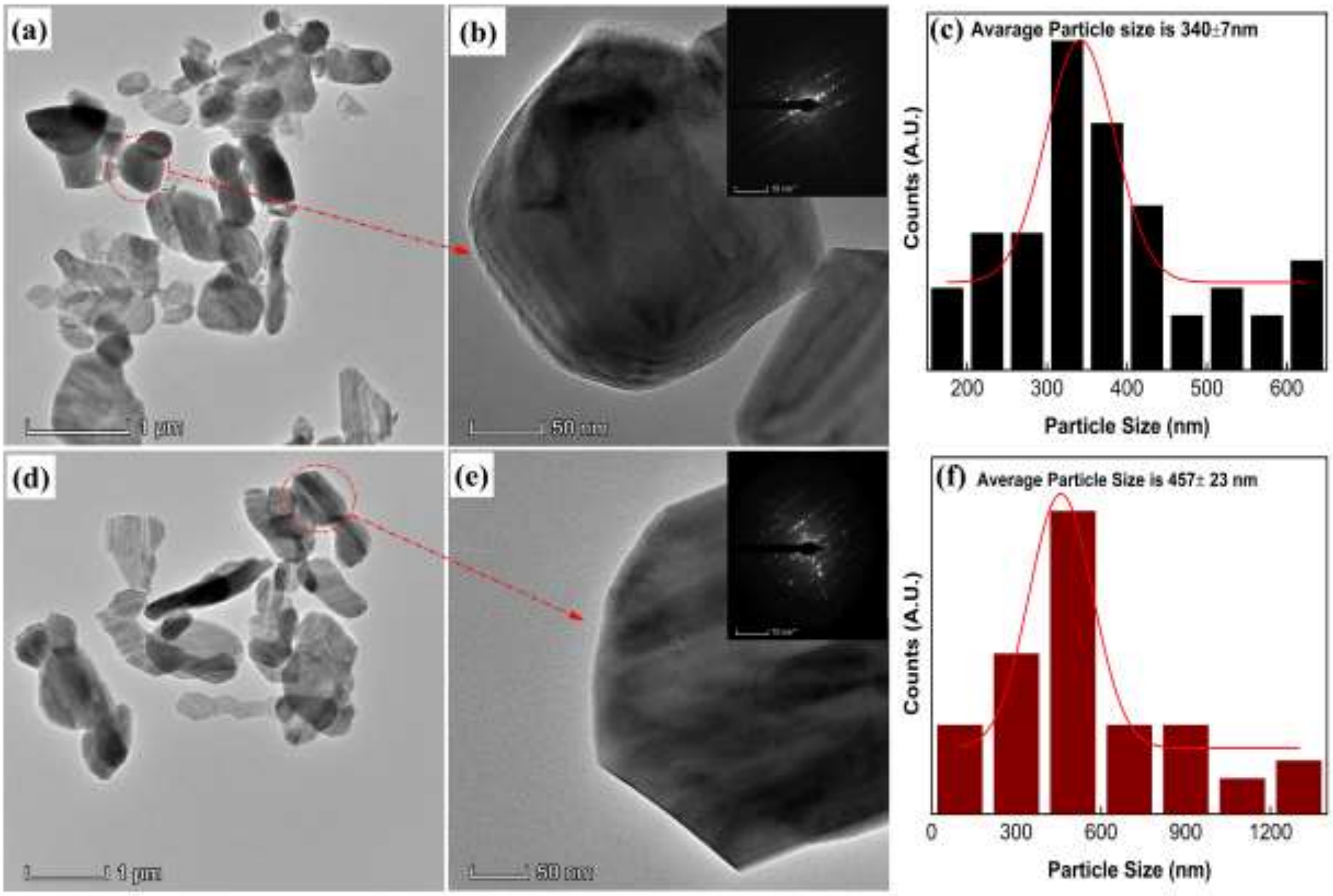

Fig. 4 : (a, d) and (b, e) high and low magnification TEM images of FMHTM mediated $V_2O_5$ at 100 °C and 180 °C reaction temperatures, respectively and inset-SAED pattern for the corresponding samples. (c) and (f) estimated particles size distribution calculated from the TEM images for reaction temperatures 100 °C and 180 °C, respectively.

The transmission electron microscope (TEM) was performed for two synthesized samples (reaction temperature 100 °C and 180 °C) for further confirmation of the particle-like morphology. Figure 4 (a, d) shows the low magnification and (b, e) shows the high magnification TEM images of $V_2O_5$ samples synthesized at 100 °C and 180 °C. From the illustration, it is seen that our synthesized samples have particle-like morphology that is also found in the SEM images. Figure 4 (c and f) shows the estimated particles size distribution calculated from the TEM images for 100 °C and 180 °C reaction temperatures, respectively. From the particles size distribution, it seems that the particle size became larger with the increase of reaction temperatures and are found to be 340±7 nm and 457±7 nm for the samples synthesized at 100 °C and 180 °C, respectively. The particle size increases with increasing reaction temperature is due to reduction of surface free energy that is consistent with the finding of the SEM images as well.

The single-crystal nature of the synthesized samples have also been confirmed by the inset selected area electron diffraction (SEAED) as depicted in figure 4 (b, e).

### 3.3. Functional Group Analysis

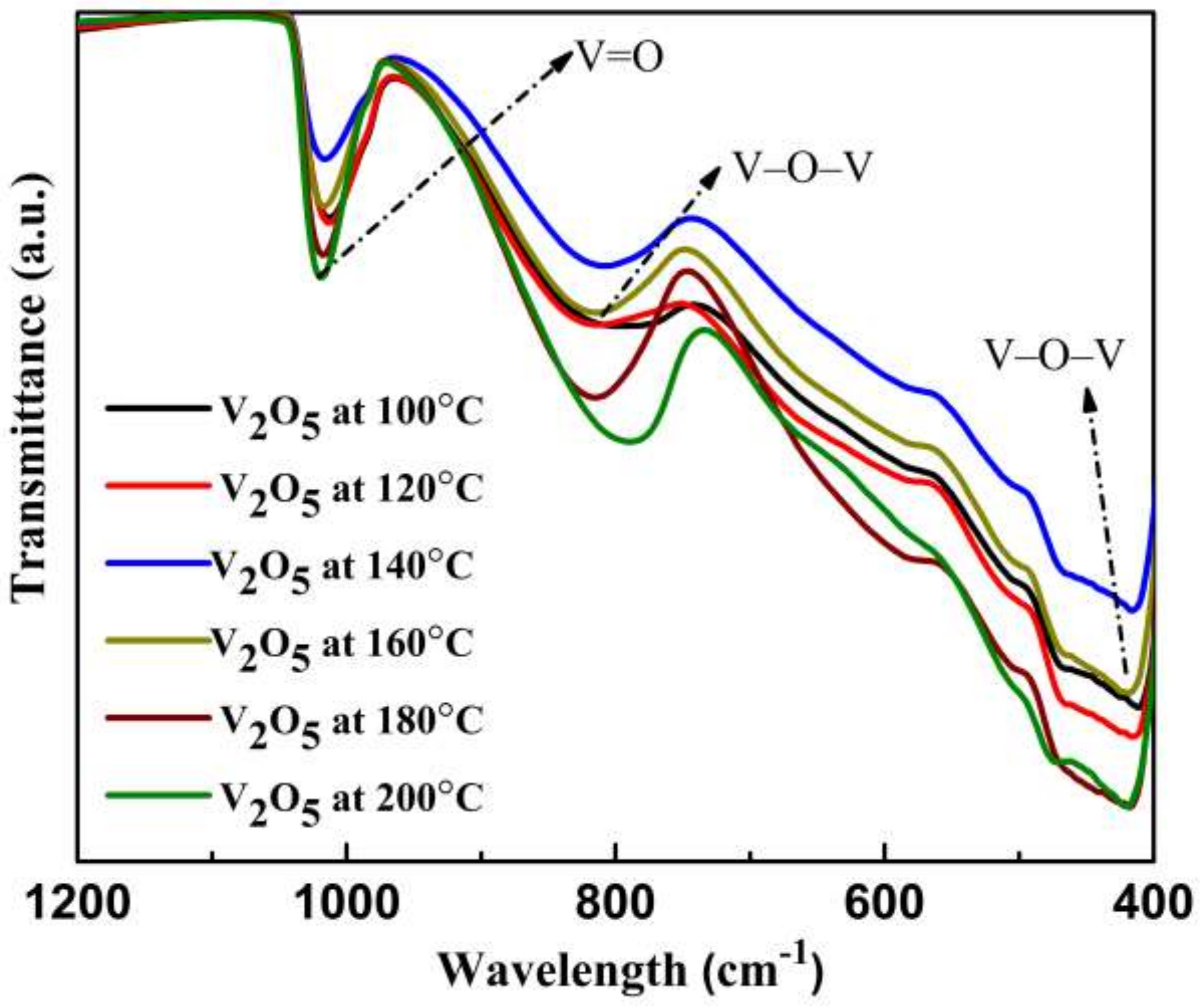

Fig. 5: The FTIR spectra of $V_2O_5$ synthesized using facile mild hydrothermal method at different reaction temperatures.

The fourier-transform infrared (FTIR) spectroscopy has been performed at room temperature with various reaction temperatures to analyze vibrational bands as illustrated in Figure 5. All three main characteristics peak of $V_2O_5$ are observed between 400 and 1000 $cm^{-1}$. First, the peak around 420 $cm^{-1}$ is for $v_s$ (symmetric stretching) modes of V-O-V vibration. It is seen that this V-O-V vibration is shifted to a lower wavelength and this downshift occurs due to wreaking of V-O-V bond [34,35]. There is a wide peak appeared around 818 $cm^{-1}$ for asymmetric stretching modes of V-O-V ($v_{as}$). The peak observed around 1020 $cm^{-1}$ is attributed to the vibrational bond for terminal oxygen (V = O). However, these two FTIR peaks confirmed the formation of $V_2O_5$ [36]. Hence, the FTIR analysis, in conjunction with the XRD analysis, affirmed the purity of $V_2O_5$.

### 3.4. Optical Properties Analysis

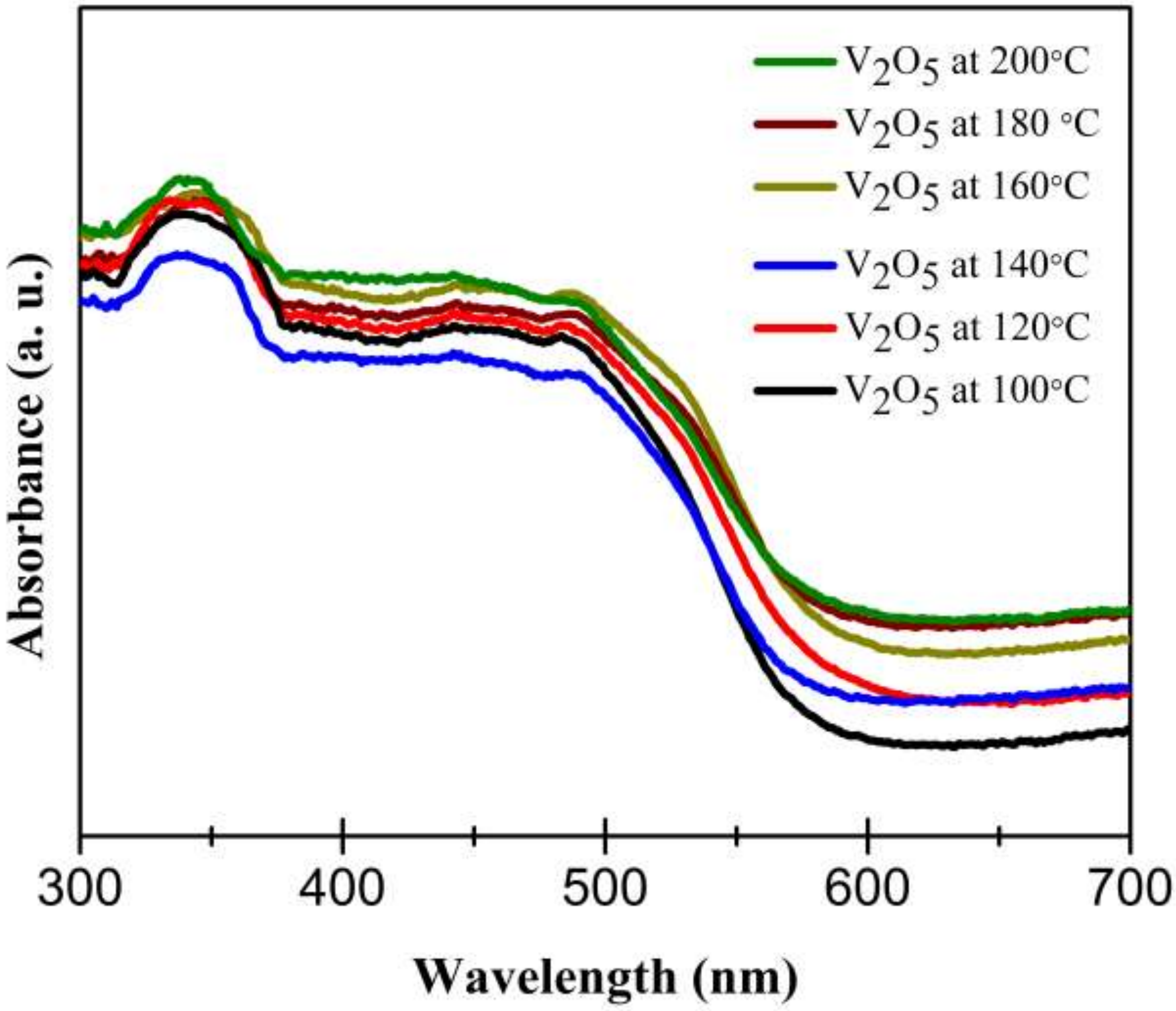

Fig. 6 Absorption spectra obtained from the UV-vis diffuse reflectance for pure $V_2O_5$ synthesized by FMHTM at different reaction temperatures.

The optical characterization was accomplished for all samples using UV-vis diffuse reflectance spectra (UV-DRS). The figure 6 illustrates the absorption spectra of pure $V_2O_5$ synthesized at six different reaction temperatures starting from 100 °C to 200 °C in step of 20°C. It is observed that all samples show a similar absorption pattern that is somewhat proven the stability of FMHTM. The $V_2O_5$ samples offer strong absorption both in the UV and visible regions. This pattern is the responsible for transferring the charge from O-2p to V-3d orbitals. Rajeshwari S. et al. also found the same absorption spectra patterns for the $V_2O_5$ synthesis [9].

The Kubelka-Munk (K-M) [37] method has been used to determine the bandgap of the pristine $V_2O_5$ powder samples. The figure 7 depicts the estimated band gap of the samples from UV-vis DRS data measured at room temperature. The relation between reflectance, R and the K-M function, F(R) according to the K-M equation are represented by the following equation[38]:

$$F(R) = \frac{(1-R)^2}{2R} \ldots\ldots\ldots\ldots\ldots\ldots\ldots\ldots\ldots\ldots\ldots\ldots 1$$

However, F(R) is directly proportional to the coefficient of absorption, $\alpha$ and using this relation $E_g$ can be obtained by plotting $[F(R)*h\upsilon]^2$ vs hυ graph. Now the relation among F(R) and $\alpha$ is given by the equation,

$$F \propto \alpha$$

$$F(R) * hv = A\left(hv - E_g\right)^n \ldots\ldots\ldots\ldots\ldots\ldots\ldots .2$$

Here in the above equation, $h\upsilon$ is the energy of photon, $h$ for Plank's constant, $A$ is the proportionality constant and $E_g$ is for the desired bandgap energy. In this equation, n, determines whether the optical band gap is direct or indirect.

Table I: Calculated values of $E_g$ for synthesized samples

| Sample Name | Hydrothermal Reaction Temperature (°C) | Calcination temperature | $E_g$ (eV) |
|---|---|---|---|
| $V_2O_5$ | 100 | 500°C | 2.08 |
| | 120 | | 2.10 |
| | 140 | | 2.09 |
| | 160 | | 2.11 |
| | 180 | | 2.11 |
| | 200 | | 2.15 |

The values of n are used ½ and 2 for direct and indirect optical band gap, respectively. Here, we used n=2 for calculating the direct bandgap of all synthesized samples.

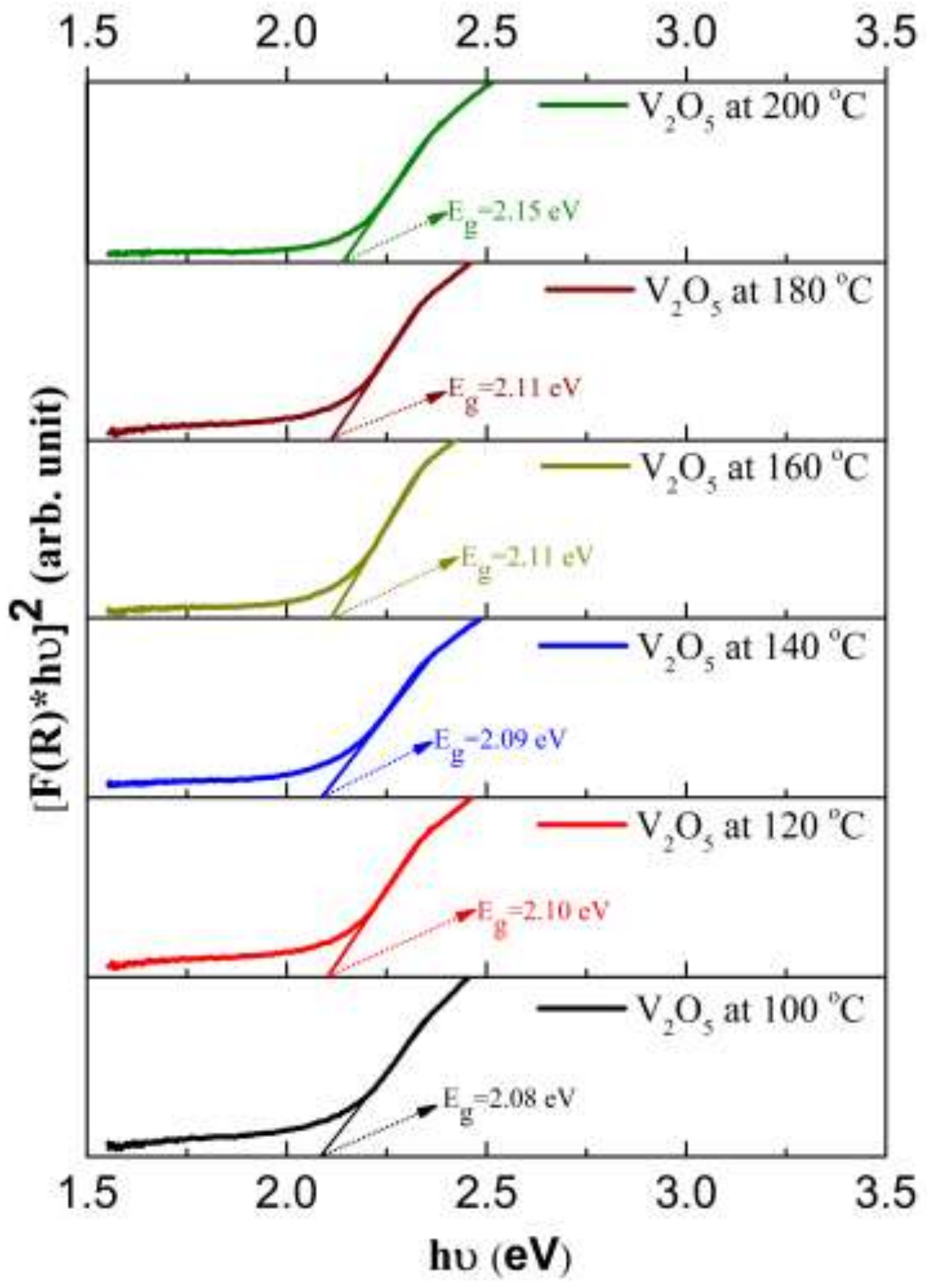

Fig. 7 The $[F(R)*h\upsilon]^2$ vs hυ plot to estimate the band gap of synthesized $V_2O_5$ samples.

The estimated bandgap at room temperature of the synthesized samples are summarized in Table I. It is seen that the $V_2O_5$ synthesized at 100°C exhibits lower bandgap (2.08 eV) then gradually increased to 2.15 eV for the $V_2O_5$ synthesized at 200°C. This rising trend can be explained using familiar concept of Burstein-Moss (B-M) for the semiconductor materials where particle size rises with the increase of reaction temperature results the Fermi level moves to the conduction band hence bandgap increases [39]. However, all samples can be able to absorb a significant amount of visible light as their bandgap varies from 2.08 eV to 2.15 eV.

## 3.5. Photocatalytic Degradation Activity Analysis

Photocatalytic degradation activity of methylene blue (MB) dye using two $V_2O_5$ samples, one synthesized at low reaction temperature (100°C) and another at high reaction temperature (180°C), The photocatalysis have been performed in aquatic medium (using deionized water) irradiated by visible light (LED) for up to 180 min. Figure 8 (a) and (b) illustrate obtained time variable UV-vis absorption spectra of $V_2O_5$ synthesized at 100°C and 180°C, respectively. Figure 8 (c) represents the $C/C_0$ plot for the MB degradation for both samples where $C_0$ represent the concentration of the MB at adsorption equilibrium, and C represents the concentration of dye remaining non-degraded after the irradiation done for a specific time. The amount of degradation efficiency is estimated using the following formula,

$$\text{Degradation (\%)} = (1 - C/C_0) \times 100\% \quad \text{………………………3}$$

The $\ln(C_0/C)$ versus time (t) plot for the MB dye is represented in Figure 8 (d). The rate constants are determined using the equation $\ln(C_0/C) = kt$, where $k$ is the degradation rate constant.

From figure 8 (c) it is seen that the photodegradation efficiencies are 40% and 25% within 180 min of irradiation for the $V_2O_5$ synthesized at 100°C and 180°C, respectively. So, $V_2O_5$ synthesized at lower reaction temperature shows higher efficiency. The rate constant ($k$) calculated from the linear fitting of logarithmic plot MB dye are $(2.0 \pm 0.1) \times 10^{-3}$ $min^{-1}$ and $(1.7 \pm 0.1) \times 10^{-3}$ $min^{-1}$ for the $V_2O_5$ synthesized at 100°C and 180°C, respectively. This higher $k$ value also confirmed that the $V_2O_5$ synthesized at 100°C has higher efficiency. We anticipate this fact is that the smaller particle size, and accordingly, the higher surface to volume ratio and also lower bandgap of $V_2O_5$ synthesized at 100°C than that of $V_2O_5$ synthesized at 180°C are the reason behind the higher efficiency [40]. However, the photocatalytic result is not good enough. To us, this might be due to the presence of various types of defects, which acts as rapid recombination, thus reducing the generation of active species.

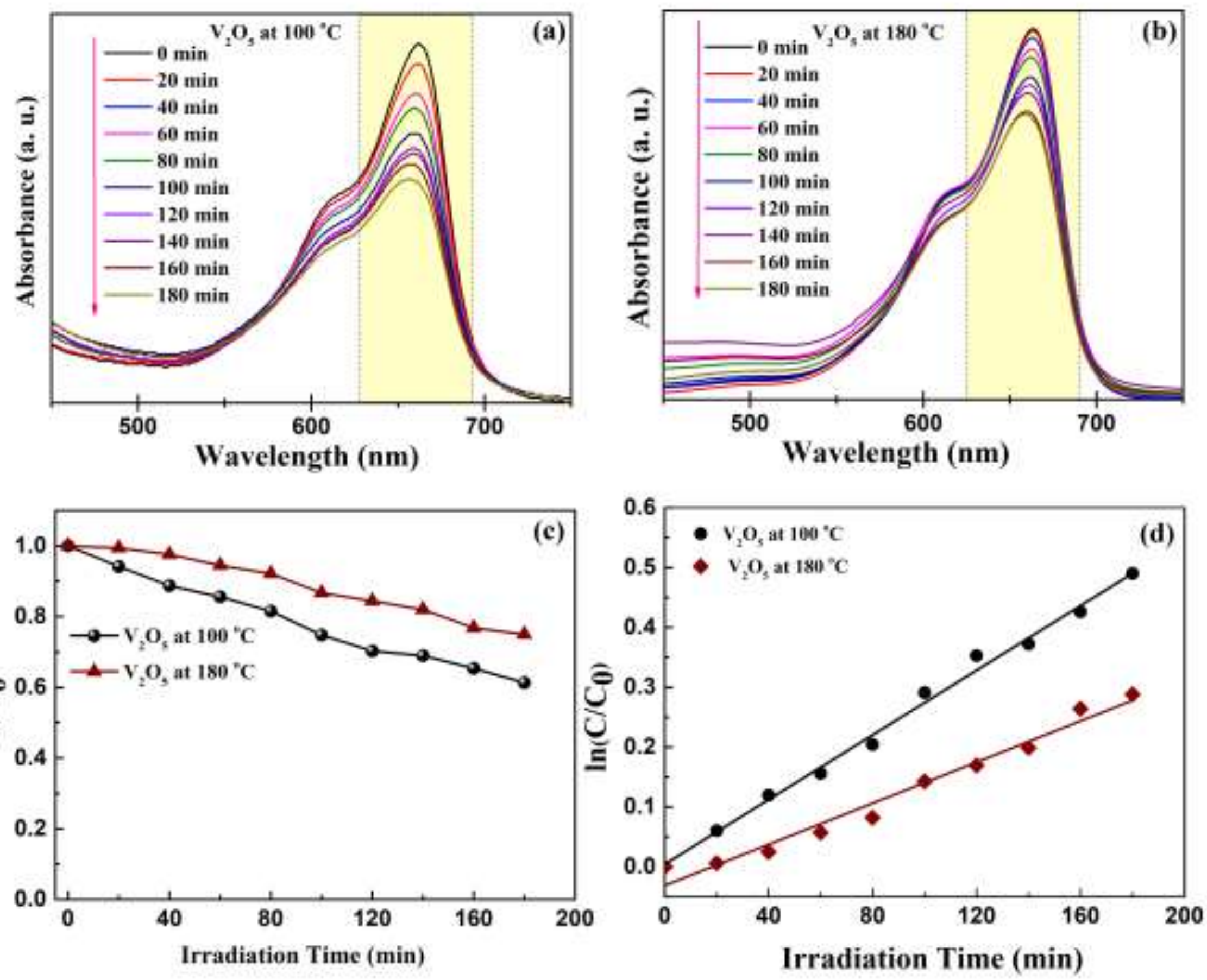

Fig.8 The photocatalytic illustration. (a, b) time-variable absorption spectra, (c) photodegradation efficiency of $V_2O_5$, and (c) the $\ln(C_0/C)$ vs Irradiation time graph

## Photodegradation mechanism of $V_2O_5$

In a typical photodegradation, when the electromagnetic radiation falls on a photocatalyst then the photons are absorbed, and the electrons in the valence band are excited. The excited electrons are then migrate to the conduction band and, accordingly, the electron-hole produces. It should be noted that the radiation energy should be equal to or greater than that of the bandgap of the photocatalyst. The electron-hole is then produce radicals, and these radicals are participated in the photodegradation process of the contaminants.

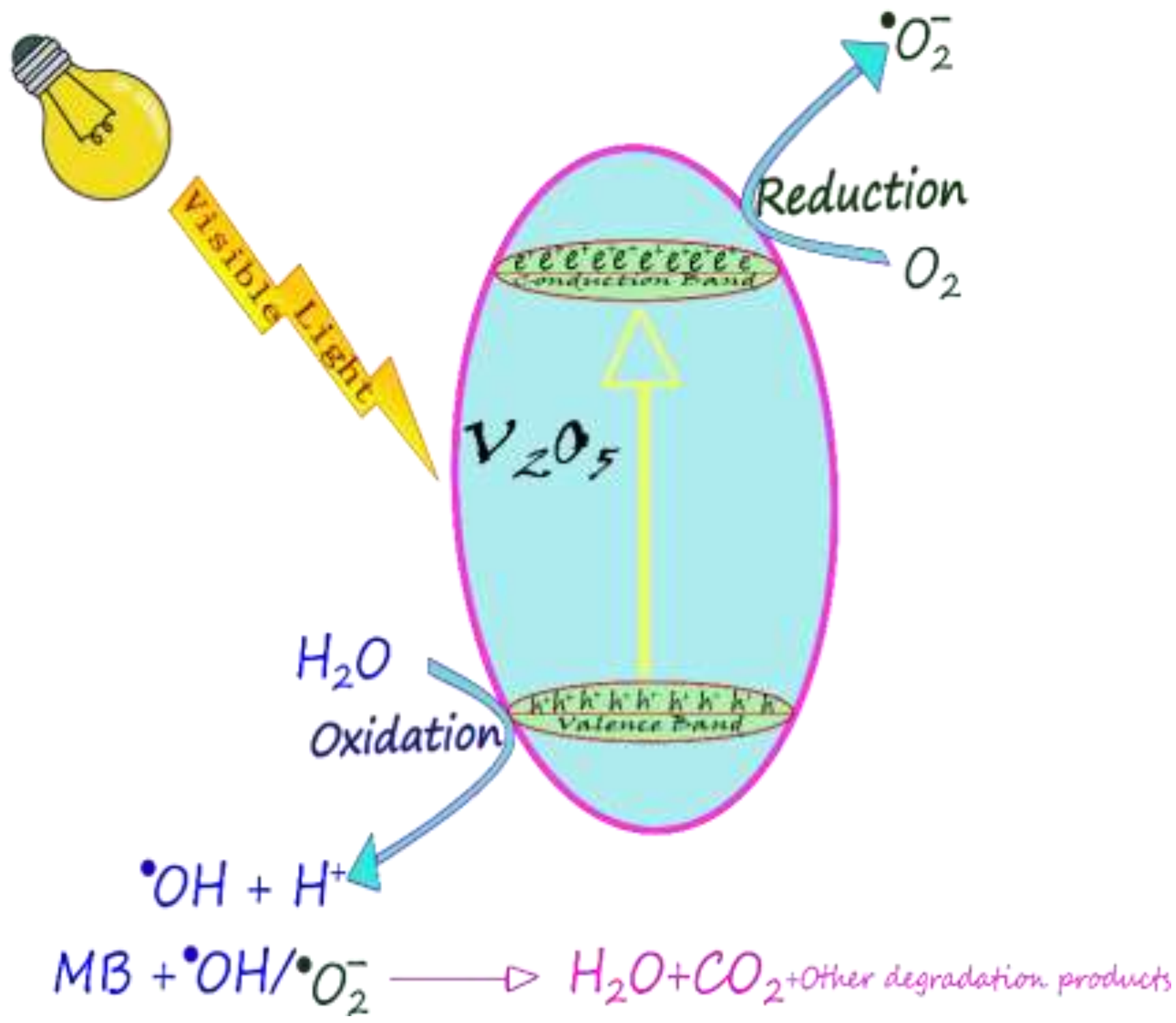

Fig. 9: Schematic illustration of the photocatalytic mechanism.

However, there are several redox reactions involved in the degradation of MB under the irradiation of light. First, when the photon of suitable energy interacted with the $V_2O_5$ then the electron-hole pair is produced (Equation 4).

$$\text{Photocatalyst}, V_2O_5 + \text{light}\ (h\nu) \longrightarrow h_{VB}^+ + e_{CB}^- \ldots\ldots\ldots\ldots\ldots\ldots 4$$

$$Reduction: \quad e_{CB}^- + O_2 \longrightarrow {}^{\circ}O_2^-\ (Superoxide\ radical) \ldots\ldots\ldots\ldots\ldots 5$$

$$H_2O \longrightarrow OH^- + H^+ \ldots\ldots\ldots\ldots\ldots\ldots\ldots 6$$

$$Oxidation: \quad h_{VB}^+ + OH^+ \longrightarrow {}^{\circ}OH\ (Hydroxyl\ radical) \ldots\ldots\ldots\ldots\ldots 7$$

$$Degredation: Dye(MB) + {}^{\circ}OH\ \ or\ {}^{\circ}O_2^- \longrightarrow CO_2 + H_2O \ldots\ldots\ldots\ldots\ldots 8$$

Then, the electrons, $e_{CB}^{-}$ from conduction band interacts with the $O_2$ to produce superoxide radicals, $°O_2^{-}$ (Equation 5, Reduction). The holes, $h_{VB}^{+}$from the valence band interacts with $OH^{-}$of the water to form hydroxyl radicals, $OH^{+}$ (Equation 7, Oxidation,). Finally, both superoxide and hydroxyl radicals ($°O_2^{-}$ $and$ $OH^{+}$) participate in the degradation of dye, MB (Equation 8).

## 4. CONCLUSION

We have successfully synthesized $V_2O_5$ from ammonium metavanadate at six different reaction temperatures (100°C to 200°C, in the step of 20°C) using an environment-friendly facile mild hydrothermal method. The stable orthorhombic structure of $V_2O_5$ was confirmed by the XRD analysis. The SEM and TEM study shows that all samples hold particle-like morphology, and reveal that the particle size increases with the increase of reaction temperatures. Furthermore, the SEAED images affirmed the single-crystal nature of the synthesized samples. The FTIR study further confirmed the formation of single phase $V_2O_5$. The optical bandgap increases with the rise in reaction temperatures and, accordingly, exhibits the lower photocatalytic efficiency. The degradation efficiency is found to be 25% for 180°C reaction temperature whereas 40% for 100°C. However, the degradation efficiency is not judicious though the bandgap is suitable for higher photocatalytic efficiency. The lower photocatalytic efficiency can be explained due to the presence of various defects that acts as center of rapid recombination results diminishes the generation of active species accordingly declines the photocatalytic efficiency. It is noteworthy that tuning the calcination temperature can further reduces particles size and expect to increase the surface to volume ratio and hence efficiency.

**Acknowledgements:**

The authors are grateful to the Directorate of Research and Extension (DRE), Chittagong University of Engineering and Technology (CUET), Chattogram-4349, Bangladesh and University of Grant Commission (UGC) for arranging financial assistance under grant numbers CUET DRE (CUET/DRE/2018-19/PHY/008) and ----------------- for arranging the financial support for this work. We are also thankful for the laboratory support of the Materials Science Division, Atomic Energy Centre, Dhaka 1000, Bangladesh for the experimental support.